\documentclass[aps,twocolumn,showpacs,superscriptaddress]{revtex4}

\usepackage[dvips]{graphicx}

\begin{document}

\newcommand{\be}{\begin{equation}}
\newcommand{\ee}{\end{equation}}
\newcommand{\ba}{\begin{eqnarray}}
\newcommand{\ea}{\end{eqnarray}}
\newcommand{\tr}{{\rm tr}}
\newcommand{\bea}{\begin{eqnarray}}
\newcommand{\eea}{  \end{eqnarray}}
\newcommand{\cnot}{\textsc{cnot}}

\title{Quantum baker maps with controlled-NOT coupling}

\author{Ra\'ul O. Vallejos}
\email{vallejos@cbpf.br}
\homepage{http://www.cbpf.br/~vallejos}
\author{Pedro R. del Santoro}
\email{psantoro@ifi.unicamp.br}
\author{Alfredo M. Ozorio de Almeida}
\email{ozorio@cbpf.br}
\homepage{http://www.cbpf.br/~ozorio}
\affiliation{Centro Brasileiro de Pesquisas F\'{\i}sicas (CBPF), 
             Rua Dr.~Xavier Sigaud 150, 
             22290-180 Rio de Janeiro, 
             Brazil}

\date{\today}

\begin{abstract}
The characteristic stretching and squeezing of chaotic motion is linearized
within the finite number of phase space domains which subdivide a classical 
baker map.
Tensor products of such maps are also chaotic, but a more interesting
generalized baker map arises if the stacking orders for the factor maps
are allowed to interact. 
These maps are readily quantized, in such a way that the stacking interaction 
is entirely attributed to primary qubits in each map, if each j'th subsystem 
has Hilbert space dimension $D_j=2^{n_j}$.  
We here study the particular example of two baker maps that interact via a 
{\it controlled-not} interaction, which is a universal gate for quantum 
computation.
Numerical evidence indicates that the control subspace becomes an
ideal Markovian environment for the target map in the limit of large Hilbert 
space dimension.
\end{abstract}



\maketitle

\section{Introduction}

The baker map displays the essential features of classically chaotic 
motion in such a simplified form as to be almost a caricature. 
It may be described as a space-filling horseshoe map that linearizes 
the hyperbolic motion. 
Thus, essentially chaotic motion is exhibited, which can be followed 
through very simple computations.
The binary symbolic dynamics propagates vertical strips 
in the unit square onto horizontal strips. 
The vertical and horizontal rectangles are narrower, 
for longer binary codes, 
so that the primary digit specifies a half-square.
Hence this digit is responsible for a coarse-grained 
description of the motion.
The way in which horizontal strips are piled up by the 
mapping is also determined by the primary binary digit. 

The various quantization schemes that have been proposed for the baker map 
generally respect the main division of the square into a pair of vertical 
rectangles, 
which split the position states into Hilbert subspaces. 
These are mapped respectively onto a pair of momentum subspaces according 
to the primary binary digit. 
The discrete and finite nature of the Hilbert space prevents the association 
of longer codes 
to ever thinner strips \cite{SAV}. Thus, to some extent, 
the primary digit is even more important in quantum mechanics. 
Initially, it may have seemed that the introduction of binary symbols 
to describe evolving quantum systems could only serve as an artificial prop 
for the study of the semiclassical limit. 
However, the growing interest in quantum computation 
has brought the binary structure to the fore. 
Schack noted that the quantum baker could be efficiently 
realized in terms of quantum gates \cite{schack98}.
A three qubit Nuclear Magnetic Resonance experiment was 
proposed \cite{brun99} and then implemented (with some 
simplifications) \cite{weinstein02}.
In this context, 
the quantum baker map has reemerged as an ideal simplified model. 
Schack and Caves \cite{SchackCaves} have indeed developed
an entire class of quantum baker maps, based on the $2^N$-dimensional 
Hilbert space of $N$ qubits, 
which include the original quantizations of Balazs and Voros 
\cite{BalazsVoros} 
and Saraceno \cite{Saraceno} as special cases. 
While incorporating the qubit structure and hence a Hilbert space of 
$D=2^N$ dimensions, 
we here adopt the Balazs-Voros-Saraceno map throughout, 
since this singles out the primary bit most clearly.

The object of our study is the coupling of quantum baker maps, 
initially restricted to a pair of maps.
Clearly, the higher dimensional product-map of classical baker maps 
is also chaotic, 
in spite of possible added  symmetries. The minimal coupling that 
can be meaningfully addressed 
relies exclusively on the primary binary digit.
Thus, each baker map is only sensitive to the most coarse-grained 
information
available about the state of the other map. 
In other words, one can investigate the class of coupled baker maps 
that interact through the order in which the pair of horizontal 
rectangles are piled up. 
Once these coupled maps are quantized, this corresponds to an 
interaction among only the primary qubits.

It is important at this stage to distinguish two alternative forms of 
classical-quantum correspondence for individual
baker maps. In the original quantizations \cite{BalazsVoros, Saraceno}, 
the classical phase space is two-dimensional and it is the chaotic 
motion in this two-dimensional torus that corresponds to the quantum 
level repulsion, in keeping with 
the spectra of random matrices \cite{Bohigas}. 
If the quantization is chosen with a value of Planck's constant such 
that the Hilbert space has exactly $2^N$ states, then it is possible 
to reinterpret the quantum system to be the tensor product of $N$ 
two-level systems. This would correspond to a hypercube in a 
$2N$-dimensional phase space, but it would be stretching it to 
ascribe a classical correspondence to this quintessentially quantum 
system. We shall
here keep to the original quantization schemes and, hence, interpret 
each quantum baker map as corresponding to a 
classical two-dimensional phase space, so that the coupling of two 
baker maps corresponds to a four-dimensional
phase space. 
The four possible combinations of primary binary digits 
--$(0,0), (0,1), (1,0), (1,1)$-- then correspond to
four squares in the  classical position space $(q_1, q_2)$ and hence 
to four parallelepipeds in the unit 4-cube
$(q_1, q_2, p_1, p_2)$ in which the classical motion is defined.   

Perhaps the simplest choice for coupling two classical bakers is a 
simple switch of the piling: 
Each map stacks in accordance to the 
other's primary digit. 
Obviously, there is then no change if their 
digits coincide, but, otherwise, both maps invert their piling. 
This is analogous to a mutual spin flip for the principal qubits 
of the quantized ``switch baker map". Notice that this interaction 
does not break the exchange symmetry of the product map.

As a direct model for (quantum) computing, we here concentrate on the 
``controlled-not baker map". 
In this unsymmetric case, the control map evolves just as an ordinary 
baker map, while it determines the stacking of the target map. 
The latter piles its rectangles as a normal baker when the primary 
digit of the control map is zero, but this is switched when the control 
digit is one. It is well known that the controlled-not gate is a basic 
element of quantum computing \cite{Nielsen}. The present may be considered as an 
interesting example of a system where only a singe element of a 
higher dimensional Hilbert space determines a qubit. 
This has been advocated as possibly beneficial for the stabilization 
of quantum computation against decoherence \cite{OM}.

It is possible to study the evolution of the entanglement of the 
various qubits resulting from the iteration of
an individual quantum baker map \cite{ScottCaves}. 
In the case of coupled baker maps, we can interpret the 
primary bit of each component baker map as implicated in 
quantum computation, while
the remaining qubits in each map model the local environment, 
leading to a loss of quantum coherence. Alternatively,
we here consider a second baker map as a model for the environment 
coupled to the map that is singled out as the open quantum system. 

The presentation of the class of minimally coupled maps is preceeded by a 
brief review of the ordinary two-dimensional baker map and its quantization 
in section 2. Section 3 then presents the coupled maps and their quantization 
with special emphasis on the controlled-not baker map. 
The evolution of bipartite entanglement of a product pure state is studied 
in section 4 by considering the control and the target of the controlled-not 
baker as the separate subsystems. 
The asymmetric nature of this interaction suggests that this should be an 
ideal Markovian system, in the appropriate limit, as far as the target 
is concerned. This is verified by the evolution of the linear entropy. 
Our results are discussed in section 5.

\section{Review of classical and quantum baker maps}

In this section we present the well known classical and quantum
ingredients of baker maps. However, the alternative pilings,
which are usually equivalent, need to be made explicit when
different maps are coupled.

The classical baker transformation is an area preserving,
piecewise--linear map, 
${\bf b}: (p_0,q_0) \rightarrow (p_1,q_1)$ 
of the unit square 
(periodic boundary conditions are assumed) defined as
\be
p_{1}=\frac{1}{2}(p_{0}+\epsilon_0) ~,~~~ 
q_{1}= 2q_{0}-\epsilon_0 ~;
\label{bakclas}
\ee
where $\epsilon_{0}=[2q_{0}]$, the integer part of $2q_{0}$.  This map
is known to be uniformly hyperbolic, the stability exponent for orbits
of period $L$ being $L\log 2$.  Moreover it admits a useful description
in terms of a complete symbolic dynamics. A one to one correspondence
between phase space coordinates and  binary sequences,
\be
(p,q) \leftrightarrow 
\ldots 
\epsilon_{-2} 
\epsilon_{-1} 
\cdot 
\epsilon_{0} 
\epsilon_{1} 
\epsilon_{2} 
\ldots
~~~,~ \epsilon_i=0,1 ~,
\label{symbol}
\ee
can be constructed in such a way that the action of the map is
conjugated to a shift map.  The symbols are assigned as follows:
$\epsilon_i$ is set to zero (one) when the $i$--th iteration of 
$(p,q)$ falls to the left (right) of the line $q=1/2$, i.e. 
$[2q_i]=\epsilon_i$. 
Reciprocally, given an itinerary
\be
\ldots 
\epsilon_{-2} 
\epsilon_{-1} 
\cdot 
\epsilon_{0} 
\epsilon_{1} 
\epsilon_{2} 
\ldots,
\label{symbolic}
\ee 
the related phase point is obtained through the specially
simple binary expansions
\be
q=\sum_{i=0}^{\infty} \frac{\epsilon_i}{2^{i+1}} ~,~~~
p=\sum_{i=1}^{\infty} \frac{\epsilon_{-i}}{2^i} ~.
\label{pqsym}
\ee
Once the dynamics has been mapped to a shift on binary sequences it is
very easy to analyze the dynamical features of the map.  In particular,
periodic points are associated  to infinite repetitions of {\em finite}
sequences of symbols.

Due to its piecewise linear nature, the baker map admits a (mixed)
generating function which is a piecewise bilinear form,
\be
W_{\epsilon_{0}}(p_{1},q_{0})=
2p_{1}q_{0} - \epsilon_{0}p_{1} -\epsilon_{0}q_{0} 
~,~~~ \epsilon_{0}=0,1 ~.
\label{genf}
\ee
It is not defined on the whole space $(p_1,q_0)$ but on the classically
allowed domains
\be
R_0=[0,1/2]\otimes[0,1/2] ~~~ \mbox{and} ~~~ R_1=[1/2,1]\otimes[1/2,1] ~.
\label{domains}
\ee
Though the above generating function will be the starting point for 
quantization, 
it must be remembered that it only provides an implicit formula for
each iteration of the classical baker map. 
 
With respect to the quantum map, we will follow the original
quantization of Balazs and Voros \cite{BalazsVoros}, as later modified by Saraceno
\cite{Saraceno} to preserve in the quantum map all the symmetries of its
classical counterpart. In the mixed representation the baker's
propagator can be written as a $D \times D$ block--matrix ($D$ even):
\be
\langle p_{m}|{\widehat B}_D|q_{n} \rangle= 
                \left( \begin{array}{cc}
                           G_{D/2} &      0            \\
                              0    & G_{D/2}
                       \end{array}
                \right) ~,
\label{qbdef}
\ee
where position and momentum eigenvalues run on a discrete mesh with
step $1/D=h$ ($h$ = Planck's constant), so that
\be
q_{n}=(n+1/2)/D ~,~~~ p_{m}=(m+1/2)/D ~,~~~ 0 \le n,m \le D-1 ~;
\label{grid}
\ee
and $G_D$ is the antiperiodic Fourier matrix, which transforms 
from the $q$ to the $p$ basis,
\be
G_{D}=\langle p_{m}|q_{n} \rangle=(1/\sqrt D)e^{-2\pi i D p_m q_n} ~.
\label{Fourier}
\ee
It will be useful to consider this matrix as the position representation 
of the Fourier operator $\widehat{G}_D$, such that 
$\widehat{G}_D |p_{n} \rangle = |q_{n} \rangle$ .

The mixed propagator (\ref{qbdef}) has the standard structure of 
quantized linear symplectic maps \cite{OZS},
\be
\langle p_{m}|{\widehat B}_D|q_{n} \rangle =
\left\{ 
\begin{array}{cl}
\sqrt{2/D} \, e^{-i 2\pi D W_{0}(p_{m},q_{n})} &  
             \mbox{if} ~~ (p_{m},q_{n})\, \in \, R_0 \\
\sqrt{2/D} \, e^{-i 2\pi D W_{1}(p_{m},q_{n})} &  
             \mbox{if} ~~ (p_{m},q_{n})\, \in \, R_1 \\
0                                             &  
                                     \mbox{otherwise} ~.
\end{array}
\right.
\label{VanVleck}       
\ee
In this quantization, only those transitions are
allowed that respect the rule $[2p_m]=[2q_n]$, a reflection of the 
classical shift property.

To be able to iterate the quantum baker map, the state must be brought 
back to the position representation. 
This is achieved by an inverse Fourier transform,
so that the matrix (\ref{qbdef}) is multiplied by $G_D^{-1}$.  

Following Schack and Caves \cite{SchackCaves}, we can reinterpret 
this quantum map
as the evolution of $N$ qubits if the dimension of the Hilbert 
space satisfies $D=2^N$.
Then the position states can be defined as product states for the 
qubits in the basis,
\be
|q_{n} \rangle= |\epsilon_1 \rangle \otimes| \epsilon_2 \rangle 
\otimes...|\epsilon_N\rangle,
\ee
where $n$ has the binary expansion
\be
n=\epsilon_1...\epsilon_N=\sum_{j=1}^N\epsilon_j 2^{N-j}
\ee
and $q_n=(n+1/2)/D=0\cdot \epsilon_1...\epsilon_N 1$.
The connection with the classical baker map is specified by the symbolic 
dynamics. Finite segments of
the bi-infinite strings (\ref{symbolic}) that determine points 
in the unit square
are made to correspond to orthogonal quantum states.
Half of the position states lie in either of the two rectangles, $R_0$, 
or $R_1$ defined in (\ref{domains}), which correspond respectively 
to $0$, 
or $1$ eigenstates of the principal qubit.

The full unitary operator for the quantum baker 
can be written explicitly as
\be 
{\widehat B}_D={\widehat G}_D[{\hat 1}_2 \otimes {\widehat G}^{-1}_{D/2}],
\label{boperator}
\ee 
where ${\hat 1}_2$ is the unit operator for the first qubit and 
${\widehat G}^{-1}_{D/2}$
is the inverse Fourier operator on the remaining qubits. 
The operator in the square brackets
preserves the first qubit, while evolving separately the remaining qubits,
within each domain $R_{\epsilon_1}$. It is the final
Fourier operator that mixes the principal qubit in with the rest,
because it acts globally on the states in both domains. 
This step is not explicit in the mixed representation (\ref{qbdef}).  

So far we have only allowed for a single possibility in which to stack
the rectangles in the baker transformation, but alternative to (\ref{bakclas}),
the classical map ${\bf b'}:(p_0,q_0)\rightarrow(p_1,q_1)$,
\be
p_{1}=\frac{1}{2}(p_{0}+1-\epsilon_0) ~,~~~ 
q_{1}= 2q_{0}-\epsilon_0 ~;
\label{bakclas'}
\ee
has very similar properties. This variation is tantamount to
reversing the primary classical bit and it corresponds to the quantum map
$\widehat {B}'$, represented by the matrix:
\be
\langle p_{m}|{\widehat B'}_D|q_{n} \rangle= 
                \left( \begin{array}{cc}
                           0 & G_{D/2}            \\
                     G_{D/2} & 0
                       \end{array}
                \right) ~.
\label{qbdef'}
\ee
In other words, we here substitute the operator ${\hat I}_2$, which acted
on the principal qubit, by ${\widehat X}_2$ represented by the Pauli matrix:
\be
X_2=\left( \begin{array}{cc}
                           0 & 1            \\
                           1 & 0
                       \end{array}
                \right) ~,
\ee
so that ${\widehat B'}_D={\widehat G}_D[{\widehat X}_2 \otimes 
{\widehat G}^{-1}_{D/2}]$.

Of course, we are free to substitute any other unitary operator acting on
the primary qubit, but it is only $\widehat {B}'$ that can be interpreted
classically as an equivalent alternative piling of the baker map.
We can still split up the evolution into domains equivalent to (\ref{domains}),
though with a different matching of $q_0$ and $p_1$ segments. 
The classical evolution within each domain is again determined by classical
generating functions like (\ref{genf}) which become the exponent of the 
propagator.

If we allow a finite probability for a stacking fault in the classical
baker map, the evolution acquires a random component. However,
if the piling order depends on the coarse-grained position of another
baker map, the overall motion will again be purely deterministic.
Being that each baker map is thoroughly chaotic, it will be hard to
distinguish the random motion of one of the components taken on its own
from that of a chaotic system with an added stochastic variable.
In the following section we allow the baker to choose between these
alternatives, depending on its interaction with another baker.

\section{Coupling by the primary qubit}

All possible couplings of a pair of quantum qubits are specified
by unitary matrices acting on the basis states $(00, 01, 10, 11)$.
In particular,
\be
{I}_2 \otimes {I}_2= \left( \begin{array}{cccc}
1 & 0 & 0 & 0   \\
0 & 1 & 0 & 0   \\
0 & 0 & 1 & 0   \\
0 & 0 & 0 & 1 
\end{array}  \right) ~,
\ee
leaves both qubits invariant. This is the correct description
for the pair of principal qubits for two uncoupled quantum baker maps,
$\widehat B \otimes \widehat B$. Likewise, a pair of uncoupled
bakers, $\widehat {B}' \otimes \widehat {B}'$, 
with the alternative stacking discussed in the previous section
are propagated by the matrix
\be
X_2 \otimes X_2= \left( \begin{array}{cccc}
0 & 0 & 0 & 1   \\
0 & 0 & 1 & 0   \\
0 & 1 & 0 & 0   \\
1 & 0 & 0 & 0 
\end{array}  \right) ~.
\ee

Perhaps an analogy may bring home the subtle nature of the interaction
that is now introduced between the two subsystems. 
A classical analogue for these finite numbers of
qubits could be a pair of card packs. We do not allow here any exchange of
cards between the packs, such as in a chromosome crossover. 
Each pack is always
shuffled separately by splitting in two and repiling, but each affects the 
order of the 
other's stacking. 
The interesting interactions are the ones which may,
but do not always affect the piling order of the component baker map.
One possibility is the 
\be
\textsc{swap} = \left( \begin{array}{cccc}
1 & 0 & 0 & 0   \\
0 & 0 & 1 & 0   \\
0 & 1 & 0 & 0   \\
0 & 0 & 0 & 1 
\end{array}  \right) ~.
\ee
Here the individual bakers swap their stacking, but there results no change
if they already shared the same digit, $0$ or $1$. 

Perhaps, the nicest example in the context of quantum information
is the {\it controlled-not gate}, $\widehat {\cnot}$, represented by
the unitary matrix:
\be
\cnot = 
\left( \begin{array}{cccc}
1 & 0 & 0 & 0   \\
0 & 1 & 0 & 0   \\
0 & 0 & 0 & 1   \\
0 & 0 & 1 & 0 
\end{array}  \right) ~.
\ee
This is one of the universal elements for the design of
circuits in the theory of quantum computation. Here,
the interaction goes only one way: If the corresponding classical control bit
were $0$, the target bit would not be changed, propagating the full map 
as ${\bf b}$, but this switches to ${\bf b'}$ for the control $1$.
In quantum mechanics the amplitudes for both choices are superposed,
which entangles the primary control qubit to the target qubits, so that eventually the
full set of qubits of both maps are also entangled. 

The general construction of interacting quantum baker maps is now presented
through the example of the controlled-not interaction. 
Naturally, the control qubit now becomes the control
subsystem with Hilbert space dimension $D_c$, whereas the target subsystem has
dimension $D_t$. Even if the interaction is more symmetric, $D_c$
will be the dimension of the Hilbert space 
related to the qubit that specifies $2\times 2$ blocks in the above $4\times 4$ matrices,
whereas $D_t$ is related to the qubit within each block.
The structure of the unitary operator for interacting baker maps is then
\be
{\widehat B}_\cnot = 
( {\widehat G}_{D_c} \otimes {\widehat G}_{D_t} ) \,
[ \widehat{\cnot} \otimes ({\widehat G}^{-1}_{D_c/2}
\otimes{\widehat G}^{-1}_{D_t/2})].
\label{cnotbakers}
\ee
This is an obvious generalization of the quantum baker map (\ref{boperator}).
The Fourier transforms become tensor products of Fourier transforms and
the identity for the principal qubit is substituted by $\widehat{\cnot}$ for
the pair of principal qubits. Variations within the same structure are obtained
by inserting other two qubit transformations instead of the controlled-not gate.
 
The unitary matrix that describes a single step of the coupled
maps in the mixed representation is then
\be
\langle p_{m^\prime}^c p_{m}^t
|{\widehat B}_\cnot |
        q_{n^\prime}^c q_{n}^t \rangle= 
\left( \begin{array}{cccc}
M & 0 & 0 & 0   \\
0 & M & 0 & 0   \\
0 & 0 & 0 & M   \\
0 & 0 & M & 0 
\end{array}  \right) \; ,
\label{Bcnot}
\ee
with
\be
M=G_{D_c/2}\otimes G_{D_t/2} \; .
\ee
Each of the four nonzero blocks in the matrix above is a 
$D_c/2\times D_t/2$ (inverse) Fourier transform
for the states corresponding to four classical domains,
namely products of the rectangles similar to those defined by 
(\ref{domains}). 
Within each of these 4-dimensional parallelepipeds, 
the classical evolution
is just the linear hyperbolic motion specified by generating 
functions that
are the sum of those for the baker maps {\bf b}, i.e. (\ref{genf}), 
or {\bf b'}.
Just as with a single baker map, it is the final $D_c\times D_t$ Fourier 
transformation, 
returning to the position representation, that effects an interaction 
of the principal qubits with all the others. But it should be noticed that
there is no interaction between the secondary qubits of the control 
Hilbert space
and those of the target prior to the final Fourier transformation.

The classical correspondence for the interacting quantum baker maps 
only makes sense
if both the dimensions $D_c$ and $D_t$ are large. 
However, 
it is interesting to consider the limit where the control subsystem is a single
qubit, i.e. $D_c=2$. In that case, the full map is just
\be
{\widehat B}_\cnot = ({\widehat G}_{D_c} \otimes {\widehat G}_{D_t})
[\widehat{\cnot} \otimes {\widehat G}^{-1}_{D_t/2}],
\ee
where ${\widehat G}_{D_c}$ is a single qubit gate 
(had we used the periodic Fourier transform, it would be the Hadamard gate). 
In a way, we have here removed the internal degree of freedom of the 
control, which now only functions through its effect on the target.
This is somewhat similar to the generalizations of the quantum baker map
proposed by Schack and Caves \cite{SchackCaves}, which single out more 
than a single qubit.
In the case of a pair of qubits, their map becomes 
\be 
{\widehat B}_{N,2}=
{\hat 1}_2\otimes
[{\widehat G}_{2D_t}^{-1}({\hat 1}_2\otimes{\widehat G}_{D_t})]
{\hat S},
\ee
where we have identified $D_t=2^{N-2}$, while ${\hat S}$ permutes the 
pair of principal qubits.
The resemblance becomes stronger if we substitute the $\widehat{\cnot}$ 
operator in the
interacting bakers by ${\hat 1}_2\otimes{\hat 1}_2$. 
This comparison shows that our
general construction of interacting maps 
allows for the investigation of a much richer range of dynamics 
than that displayed by previous generalizations of quantum baker maps, 
even when the internal structure is removed from one of the maps.

\begin{figure}
\includegraphics[width=9cm]{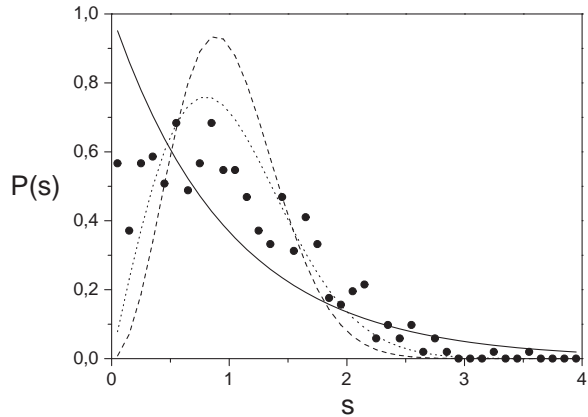}
\caption{%
Distribution of level spacings --normalized to unit average-- 
for the {\mbox \cnot} baker map (big dots). 
The dimensions are 32 and 16 for the control and target bakers, 
respectively. We also display the predictions of random matrix
theory (dotted line: GOE, dashed: GUE) and the Poisson distribution
(full line).}
\label{fig1}
\end{figure}

In spite of the fact that the quantum controlled-not baker corresponds 
to a classically chaotic map, its spectral statistics does not comply 
with one of the standard ensembles of random matrix theory \cite{Bohigas}, 
as we see in Fig.~\ref{fig1}.
This discrepancy is shared by many previous examples \cite{BalazsVoros,Sano}. 
In most cases, hidden symmetries, 
or arithmetic anomalies have been 
identified, which distinguish the system from the generic ensemble
\cite{Bogomolny}.

The classical controlled-not baker 
${\bf b}_{\cnot}:(p_0, q_0)\rightarrow(p_1, q_1)$, 
corresponding to the unitary operator (\ref{cnotbakers}), is 
\bea
p_{1}^c & = & \frac{1}{2}(p_{0}^c+\epsilon_0^c) \; ,  \\
q_{1}^c & = &  2q_{0}^c-\epsilon_0^c            \; ,  \\
p_{1}^t & = &  \frac{1}{2}
(p_{0}^t+\epsilon_0^t+\epsilon_0^c-2\epsilon_0^t\epsilon_0^c) \; , \\
q_{1}^t & = &  2q_{0}^t-\epsilon_0^t \; . 
\label{bcnot}
\eea
Thus the classical orbits depend on the binary digits, 
$\epsilon_0^c$ and $\epsilon_0^t$
in an unsymmetric way. 
The orbit of the control subsystem is independent of the behaviour 
of the target, while the evolution of the principal binary digit 
of the control, 
that results from its internal chaotic motion, affects the target. 
This influence is as random as a perfect coin toss, so the control 
can be considered as a perfect Markovian environment for the target. 

It is important to distinguish the classical trajectories
of the full 4-dimensional map, from its projections into the 
2-dimensional phase space rectangles that describe each of the 
subsystems. 
The full trajectory is deterministic, which is also the case for its 
projection onto the control phase space. 
It is the projection onto the target phase space that acquires an important 
random component, when knowledge of the control orbit is deleted. 
Viewed from the control phase space, we find an infinite number
of trajectories of the full system that project onto the same control 
orbit.  

Viewed within quantum mechanics, the relation between the subsystems 
becomes less unsymmetrical. 
Their entanglement is usually measured with reference to the reduced density 
matrix of either subsystem, that is
${\hat \rho}_c$, or ${\hat \rho}_t$, 
such as the purity, 
${\rm tr}~{\hat \rho}^2$,
or, equivalently, the linear entropy \cite{Nielsen}, 
\be
S_L = 1-\tr~{\hat \rho}^2 \; .
\ee
The effect of the entanglement is an increase of linear entropy 
from that of an initial product state, for which $S_L=0$.
These measures are necessarily the same for both subsystems, no matter 
how lopsided the interaction \cite{Nielsen}. 
Nonetheless, it is legitimate to consider the target subsystem 
of the controlled-not baker interaction as an ideal candidate for a 
quantum Markovian system. 
It is this picture of the control as forming an environment for
the target that shall be studied below, rather than the alternative 
investigation of the entanglement of the principal qubits, with local 
environments taken as the set of remaining qubits of the controlled-not 
baker. 
The latter is the line taken in \cite{ScottCaves}.

\section{Evolution of entanglement: the Markovian limit}

\begin{figure}
\includegraphics[width=9cm]{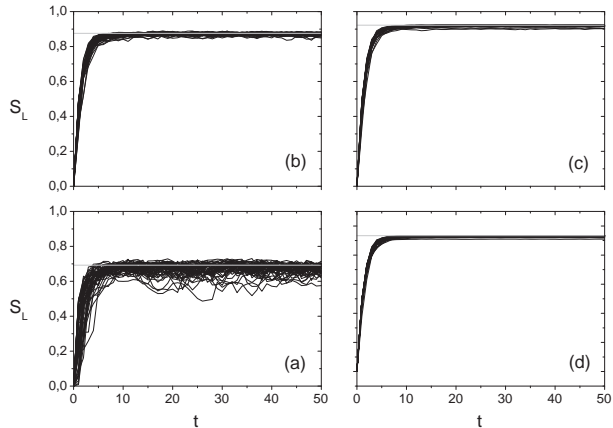}
\caption{%
Linear entropy as a function of the number of iterations of
the \mbox{\cnot} baker.
The dimension of the target space is fixed at $D_t =16$, 
but the size of the control subspace is increasing: 
$D_c$ = 4 (a), 16 (b), 64 (c), 256 (d).
In each case we considered 50 initial product states 
chosen according to the (product) Haar measure.
Also shown are the entropies of corresponding random states
(horizontal lines). }
\label{fig2}
\end{figure}

We considered the unitary evolution of the quantum controlled-not
baker map, given initial product states, 
$|\psi\rangle_c \otimes |\psi\rangle_t$. 
Both of these factor states were chosen randomly with respect to
the unitarily invariant Haar measure for pure states within 
both the control subsystem and the target subsystem \cite{Zanardi}.
Figure~\ref{fig2} show the evolution of the linear reduced entropy 
for either subsystem as a function of the discrete time, i.e. the 
number of iterations.
In each of these figures the dimension of the target space was kept 
fixed at $D_t =16$, that is, four qubits. 
However, the size of the environmental control subspace grows from 
$D_c =4$ in (a) to $D_c =256$ in (d), i.e. eight qubits.
In all cases the linear entropy rises sharply during the first 
five iterations, or so, and then oscillates around a stable average 
value, which depends on the choice of initial state.
The asymptotic values for the entropy can be compared with those
of random states \cite{ScottCaves}. 
The average value of the linear reduced entropy for the Haar 
ensemble appropriate to the full Hilbert space is less than maximal 
\cite{ScottCaves,Lubkin}:
\be
\langle S_L \rangle = 1 - \frac {D_c + D_t}{D_c D_t+1} \; .
\ee
In each of the above figures the Haar average is seen to lie 
within the fluctuations of the evolution of the individual states.
This means that the {\em entangling power} \cite{Zanardi}
of the \textsc{\cnot} baker is very close to that of a random operator
 \cite{Emerson}.
Another striking effect of increasing the dimension $D_c$
is the decrease in the amplitude of the fluctuations in 
linear entropy about their mean. 
This is in line with the decrease of the variance of the 
purity in the Haar ensemble as the dimension of the 
subsystems are increased \cite{ScottCaves}. 

\begin{figure}
\includegraphics[width=9cm]{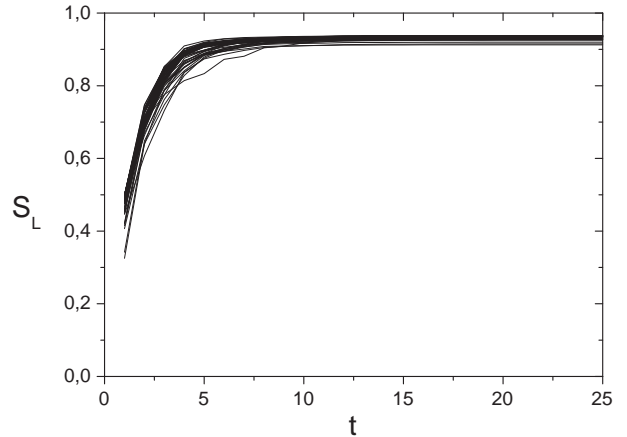}
\caption{%
Markovian limit for the target map of the \mbox{\cnot} baker.
Shown is the linear entropy as a function of the number of iterations of
the superoperator of Eq.~(\ref{Markov}).
The dimension of the target space is $D_t =16$.
We chose 50 initial product states according to the Haar measure.}
\label{fig3}
\end{figure}

Let us now compare this behaviour of the linear entropy
of the target subsystem with that of the Markovian system:
\be
{\hat \rho}_t(1) = {1 \over 2}
\left(
 \widehat B {\hat \rho}_t(0)
 \widehat B ^\dagger + 
 \widehat {B'}{\hat \rho}_t(0)
 \widehat {B'}^\dagger 
\right) \; .
\label{Markov}
\ee
That is, we average over the two possible stackings of the quantum 
baker map within the standard formalism of Kraus superoperators 
\cite{Breuer}.
This Markovian evolution is shown in Figure~\ref{fig3} for a set of
random initial target states as in Fig.~\ref{fig2}, so that for $D_t=16$. 
Evidently, the closest correspondence is verified with 
Fig.~\ref{fig2}(d), which has the largest control space. 
Even though the target only interacts directly with the principal qubit 
of the control space, the internal motion within the latter is required 
to wash away information concerning the evolution of the target.  

\begin{figure}
\includegraphics[width=9cm]{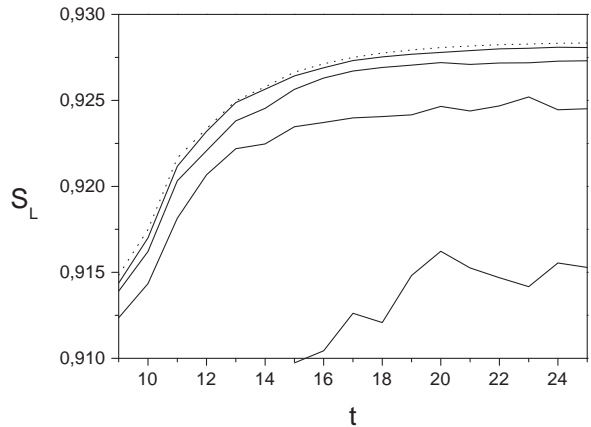}
\caption{%
Comparison of Markovian vs. unitary evolution.
(i) A single randomly-chosen initial state of the target system was evolved 
with the Markovian Eq.~(\ref{Markov}).
(ii) The same target state, tensored with the identity of the environment, 
was evolved with the full unitary dynamics.
In both cases we plot the linear entropies as a function of time.
The Markovian case corresponds to the dotted line. 
Full lines represent the unitary evolution for
environment sizes $D_t =8,16,32,64$. 
The dimension of the target space is fixed at $D_t =16$.}
\label{fig3bis}
\end{figure}

Since the Markovian approximation presupposes complete randomness
for the initial state of the environment, it is more appropriate to 
compare this irreversible evolution with the unitary evolution of a 
density operator defined as a product of a pure target state with a 
mixed control state. 
The latter is chosen to be proportional to the identity operator, 
i.e. it has equal weight for all basis state projectors in the control 
subsystem. 
Figure~\ref{fig3bis} displays the growth of the linear reduced entropy
for a single initial state chosen randomly for various dimensions of 
the control subspace, $D_c$. 
This is strong evidence that the unitary evolution of the linear 
entropy converges onto the Markovian evolution in the limit where 
$D_c \rightarrow \infty$.

\begin{figure}
\includegraphics[width=8cm]{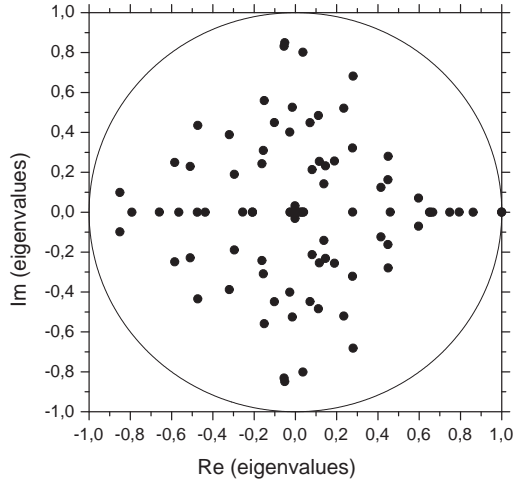}
\caption{%
Eigenvalues of the Markov superoperator.
The dimension of the target is $D_t =16$. }
\label{fig4}
\end{figure}

Most of the eigenvalues of the Markovian superoperator that
acts on ${\hat \rho}_t(0)$ in (\ref{Markov}) lie within the
unit circle, as shown in Figure~\ref{fig4}. 
In time, only the projection of the initial density operator 
with $\lambda=1$ remains.
However, this eigenvalue is doubly degenerate, corresponding
to both the identity operator and the reflection operator,
defined by $\hat R |q_m \rangle = |1-q_m\rangle$ (the reflection 
is a symmetry of the individual baker maps \cite{Saraceno}).
The fact that different initial density operators 
also differ in their projections onto this pair of states
accounts for the spread in the asymptotic values of the respective
linear entropies. 

\section{Discussion} 

The general idea of a baker map is that ``vertical" rectangular 
partitions of the unit square are stretched horizontally and 
squeezed vertically and then stacked. The generalization to 
partitions of $2L$-dimensional hypercubes is obvious, 
so that these then model a system with $L$ degrees of freedom. 
We have here shown that the different stacking orders allowed by 
higher dimensional baker maps may be interpreted as resulting from 
the various interactions of the principal qubits for each subsystem, 
if each of these is quantized to have a finite Hilbert space of 
dimension $D_j=2^{n_j}$.
As a first exploration of such systems we have chosen the 
controlled-not baker map, where the interaction of the
principal qubits is the universal \textsc{cnot} gate for quantum 
computation. 
We have discussed other possibilities for $L=2$
and there are many more for increasing $L$.

It is also possible to couple a baker map to some other simple unitary
evolution. In the case of Ermann, Paz and Saraceno \cite{EPS}, the latter
corresponds to a pair of translations. Then the control map decides the direction 
of the displacement, rather than a stacking order for the target map.
Such a quantum random walk does not belong to the class of generalized
baker maps, because the target is not a chaotic system.

A further more elaborate possibility for interaction involves a 
rotation of the four parallelepipeds into which the primary digits 
of both maps divide a four dimensional phase space. 
Choosing this to be of $\pi/4$, shuffles the parallelepipeds, 
so that the doubled binary description is respected. 
This system is studied in \cite{Santoro}. 
The interesting point is that the equilibrium and all the classical 
periodic orbits become ``loxodromic", i.e. the positions spiral outwards, 
while the momenta spiral inwards, because the eigenvalues of the stability 
matrix are general complex numbers.  
So far, not much effort has been made towards quantizing loxodromic motion,
but \cite{Rivas} is an exception. 

Unlike the case of the various quantizations of a single baker proposed by 
Scott and Caves \cite{ScottCaves}, there is a natural partition into the 
separate Hilbert spaces for interacting baker maps, with perfect classical 
correspondence.
We can analyze the entropy evolution for the reduced density 
matrix of a single subsystem for each form of coupling. 
The other clear alternative is to trace out the density matrix of 
the secondary qubits for each component
baker, so as to obtain a mixed state for each principal qubit. 
Then their entanglement is uniquely determined by the 
concurrence \cite{Wootters}. 

The remarkable facility for computing numerically the evolution of
both quantum and classical baker maps extends to their higher 
dimensional couplings.
Here we have provided numerical evidence that the generalized 
controlled-not baker map
may be an ideal model for Markovian evolution of the target map, 
once the environmental
control map is traced away. It is interesting to note that 
the interaction with the environment is not weak in any usual sense.

\acknowledgements

We are grateful to M. Saraceno for many enlightening discussions.
This research received financial support from 
CNPq, Millennium Institute for Quantum Information and PROSUL.


\begin{thebibliography}{99}

\bibitem{SAV} 
M. Saraceno and A. Voros, 
Physica D {\bf 79}, 206 (1994).

\bibitem{schack98}
   R. Schack, 
      Phys. Rev. A {\bf 57}, 1634 (1998).

\bibitem{brun99}
   A. T. Brun and R. Schack,
      Phys. Rev. A {\bf 59}, 2649 (1999)

\bibitem{weinstein02}
   Y. S. Weinstein, S. Lloyd, J. Emerson, and D. G. Cory, 
      Phys. Rev. Lett. {\bf 89}, 284102 (2002).

\bibitem{SchackCaves} 
R. Schack and C. M. Caves, 
Appl. Algebra Eng. Commun. Comput. {\bf 10}, 305 (2000).

\bibitem{BalazsVoros} 
N. L. Balazs and A. Voros, 
Ann. Phys. (NY) {\bf 190}, 1 (1989).

\bibitem{Saraceno} 
M. Saraceno M 1990,
Ann. Phys. (NY) {\bf 199}, 37 (1990).

\bibitem{ScottCaves} 
A. J. Scott and C. M. Caves,
J. Phys. A {\bf 36}, 9553 (2003).

\bibitem{Bohigas}
See for example the review by O. Bohigas in
M. J. Giannoni, A. Voros and J. Zinn-Justin (Eds.), 
Proc. Les Houches Summer School 1989, Chaos and Quantum Physics
(North-Holland, Amsterdam, 1991).

\bibitem{Nielsen}
A. Nielsen and I. Chuang,
{\it Quantum Computation and Quantum Information}
(Cambridge University Press, 2000).

\bibitem{OM}
M. C. de Oliveira and W. J. Munro
Phys. Rev. A {\bf 61}, 042309 (2000).

\bibitem{OZS} 
A. M. Ozorio de Almeida and M. Saraceno, 
Ann. Phys. (NY) {\bf 210}, 1 (1991).

\bibitem{Sano}
M. Sano, CHAOS {\bf 10}, 195 (2000).

\bibitem{Bogomolny}
E. B. Bogomolny, B. Georgeot, M.-J. Giannoni and C. Schmit,
Phys. Rep. {\bf 291}, 219 (1997).

\bibitem{Zanardi}
P. Zanardi, C. Zalka, and L. Faoro,
Phys. Rev. A, {\bf 62}, 030301 (2000).

\bibitem{Lubkin}
E. Lubkin,
J. Math. Phys. {\bf 19}, 1028 (1978).

\bibitem{Emerson}
J. Emerson, Y. S. Weinstein, M. Saraceno, S. Lloyd, and D. G. Cory,
Science {\bf 302}, 2098 (2003).

\bibitem{Breuer}
H.-P. Breuer and F. Petruccione, 
{\it Open Quantum Systems}
(Oxford University Press, 2002). 

\bibitem{EPS}
L. Ermann, J.P. Paz and M. Saraceno,
quant-ph/0510037 (2005).
 
\bibitem{Santoro}
P. R. del Santoro,
PhD Thesis (in preparation).

\bibitem{Rivas}
A. M. F. Rivas, M. Saraceno and A. M. Ozorio de Almeida
Nonlinearity {\bf 13}, 341 (2000).

\bibitem{Wootters} 
W. K. Wootters,
Phys. Rev. Lett. {\bf 80}, 2245 (1998).

\end{thebibliography}
\end{document}